\RequirePackage[l2tabu, orthodox]{nag}
\RequirePackage{snapshot}

\documentclass[10pt,onecolumn]{extarticle}

\makeatletter
\if@twocolumn
  \usepackage[dvips,letterpaper,top=0.5in, bottom=0.5in, left=0.75in, right=0.5in,includefoot,heightrounded]{geometry}
\else
  \usepackage[dvips,letterpaper,margin=1in,includefoot,heightrounded]{geometry}
\fi

\usepackage{srcltx}

\usepackage[russian,portuges,english]{babel}

\iflanguage{portuges}
    {\newcommand{\keywordname}{Palavras-chaves}}
    {\newcommand{\keywordname}{Keywords}}

\usepackage{amsmath}
\usepackage{amssymb,amsfonts}

\usepackage{abstract}

\usepackage{graphicx}
\usepackage{epsfig}
\usepackage[usenames,dvipsnames,svgnames,x11names]{xcolor}
\usepackage{subfigure}

\usepackage{booktabs}

\usepackage{setspace}
\usepackage{flushend}

\usepackage{cite}

\usepackage{hyperref}
\usepackage[normalem]{ulem}

\usepackage{enumerate}

\usepackage[noend]{algpseudocode}

\usepackage{listings}

\usepackage[short,12hr]{datetime}
       \usepackage{fouriernc}
\makeatletter

\newcommand{\printtitle}{%
\makeatletter
\if@twocolumn

\twocolumn[%
  \maketitle
  \begin{onecolabstract}
    \myabstract
  \end{onecolabstract}
  \begin{center}
    \small
    \textbf{\keywordname}
    \\\medskip
    \mykeywords
  \end{center}
  \bigskip
]
\saythanks
\else
  \maketitle
  \begin{onecolabstract}
    \myabstract
  \end{onecolabstract}
  \begin{center}
    \small
    \textbf{\keywordname}
    \\\medskip
    \mykeywords
  \end{center}
  \bigskip
  \onehalfspacing
\fi
\makeatother
}

\author{%
H. P. L. Arjuna Madanayake%
\thanks{
Department of Electrical and Computer Engineering,
University of Akron,
Akron, Ohio 44325-3904, USA}
\and
R. J. Cintra%
\thanks{%
Signal Processing Group,
Universidade Federal de Pernabuco, Brazil.
E-mail: \url{rjdsc@de.ufpe.br}}
\and
V. S. Dimitrov%
\thanks{%
Department of Electrical and Computer Engineering,
University of Calgary,
2500 University Drive N.W.,
Calgary, Alberta T2N 1N4, Canada}
\and
L. Bruton${}^\ddagger$
}

\title{%
Block-Parallel Systolic-Array Architecture for 2-D NTT-based Fragile Watermark Embedding}

\newcommand{\myabstract}{%
Number-theoretic transforms (NTTs)
have been applied in the fragile watermarking of digital images.
A block-parallel systolic-array architecture
is proposed for watermarking based on the 2-D special Hartley NTT (HNTT).
The proposed core employs two 2-D special HNTT hardware cores,
each using digital arithmetic over $\mathrm{GF}(3)$,
and processes $4\times4$ blocks of pixels in parallel every clock cycle.
Prototypes are operational on a Xilinx Sx35-10ff668 FPGA device.
The maximum estimated throughput of the FPGA circuit is
100 million $4\times4$ HNTT fragile watermarked blocks per second,
when clocked at 100~MHz.
Potential applications exist in high-traffic back-end servers dealing with
large amounts of protected digital images requiring authentication, in  remote-sensing for
high-security surveillance applications, in real-time video processing of information
of a sensitive nature or matters of national security, in video/photographic content management of
corporate clients, in authenticating multimedia for the entertainment industry, in the authentication of
electronic evidence material, and in real-time news
streaming.
}

\newcommand{\mykeywords}{%
Fragile watermarking, number-theoretic transforms
}

\date{}

\begin{document}

\printtitle

\section{Introduction}

Fragile watermarking is
an authentication technique that can detect and localize
all possible types of modifications in images including,
but not limited to, nonlinear distortion,
linear filtering,
intensity/contrast changes,
zooming,
and lossy compression of digital two-dimensional (2-D) signals.
It finds applications in
copyright protection,
high-security biometric image processing \cite{bhat1,bhat2,ret1,fprint1},
digital content management,
and Internet-based streaming video~\cite{toivonen2006video}.
These scenarios are subject to electronic attacks,
leading to an increasing demand for authentication.
Application-specific VLSI hardware for
error-free high-speed real-time fragile watermarking of large datasets
is an emerging requirement in
the information security industry~\cite{huiping2006malicious}.

In~\cite{tamori2002fragile,tamori2009asymmetric},
fragile watermarking techniques were given a number-theoretic transform (NTT) approach.
Unlike fixed-point fragile watermarking schemes based on
conventional discrete transforms~\cite{meenakshidevi2009wavelet},
systems derived from NTTs do not introduce round-off or truncation errors,
since they are capable of exact computation.
This property stems from the fact that all arithmetic computations
are performed in a finite field.
This is a highly desirable feature for practical applications~\cite{jullien1996two}.

In this work,
we propose an architecture for
the real-time computation of the
$4\times4$ \emph{special} Hartley NTT
(HNTT)~\cite{cintra2009fragile}.
The special HNTT is the finite field version of the
special discrete Hartley transform
(DHT)~\cite{watson1986separable,duleba1999medical,bracewell1986hartley}.

Furthermore,
employing the introduced 2-D special HNTT core,
an efficient architecture for
the fragile watermarking scheme by Tamori~\emph{et al.}~\cite{tamori2002fragile}
is also proposed.

It is important to note that number-theoretic transforms have
no physical meaning,
unlike the well-known conventional trigonometric transforms.
The lack of a physical meaning and concept of energy
make
the transform coefficients extremely
sensitive to error,
thereby making the NTT useful for \emph {fragile} watermaking schemes.
This high sensitivity to error is a highly-desirable property for fragile watermarking as it
guarantees complete destruction of the embedded watermark when the watermarked image is tampered-with.

The HNTT uses exact integer arithmetic,
whereas usual DHT architectures
require floating point arithmetic.

For completeness,
we provide a brief review of the available DHT architectures
based on systolic-array VLSI circuits that employ conventional DHT transforms.
Unlike the HNTT framework,
DHT architectures
suffer from finite precision effects.
Mainly represented by truncation and rounding-off,
finite precision issues
lead to performance and signal-to-noise ratios
that depend on
the employed fixed-point precisions of the systolic-array processor.

This paper unfolds as follows.
In Section~\ref{section.review},
we furnish the necessary number-theoretic results
to construct a low-complexity special HNTT.
Section~\ref{section.implementation}
offers a detailed account of the implementation
issues of the proposed hardware design.
The resulting output measurements
from the designed circuitry
are also reported
in several realistic test scenarios.
The paper is concluded in Section~\ref{section.conclusions}.

\section{Review and Mathematical Background}
\label{section.review}

\subsection{Conventional DHT Systolic-Array VLSI Architectures}

The proposed systolic-array architecture is the
\emph{first, and the only currently available}
VLSI architecture for both
the 2-D HNTT and the Tamori fragile watermarking scheme~\cite{tamori2002fragile}.
However,
we do provide
a brief summary of available VLSI architectures
so that the reader is familiar with the current state-of-the-art in
VLSI DHT architectures based on systolic-arrays although
none of the architectures reviewed here are for the HNTT.

The $N$-point
conventional DHT relates two discrete signals
$v_n$, $n=0,1,\ldots,N-1$,
and
$V_k$, $k=0,1,\ldots,N-1$,
with real components
according
to
\begin{align}
V_k
=
\sum_{n=0}^{N-1}
v_n
\mathrm{cas}
\left(
\frac{2 \pi k n}{N}
\right),
\quad
k = 0, 1, \ldots, N-1,
\end{align}
where
$\mathrm{cas}(x) = \cos(x) + \sin(x)$
is the
Hartley function.
Real-time high-speed signal and image processing algorithms
employing the DHT
achieve high-speed computations
using massively-parallel VLSI arrays.

A VLSI systolic-array design for the 1-D DHT of any length
using cyclic convolution requiring ROM based digital arithmetic
is available in~\cite{guo1994novel}.
Systolic-arrays are modular, regular, and locally interconnected,
making them very suitable for high-throughput fast algorithms.
In~\cite{BoussaktaHolt1},
a DHT is calculated using a VLSI block for
the Fermat number transform and its inverse~\cite{blahut1997fast},
at lower multiplier complexity using shift and add operations
based on the algorithm derived in~\cite{BoussaktaAlshibamiAziz1}.

In~\cite{canaris1993trigonometric},
an algorithm for calculating the DFT is extended to
a class of discrete trigonometric transforms
and
a VLSI architecture that is capable of
real time calculation of such transforms is presented.
This architecture could provide local interconnections,
identical processing elements in the systolic-array architecture
and minimal control complexity~\cite{canaris1993trigonometric}.

An application-specific VLSI architecture for
the parallel calculation of the decimation in time
and radix-2 fast Hartley transform (FHT) is available in~\cite{ZapataArg1},
where a modular and regular parallel architecture
based on a constant geometry algorithm
using butterflies of four data items and permutations is employed.
This resulted in the mapping of the algorithm to
simplified VLSI circuits and reduced communications
among processors~\cite{ZapataArg1}.

An efficient design of a VLSI systolic-array for
a prime-length type~III generalized DHT is available in~\cite{ChiperSwamyAhmad}.
The design employs an appropriate decomposition of
the generalized DHT into two half-length circular correlation structures
having the same length and form.
Such structure can be concurrently computed
and
implemented on a systolic array using hardware sharing.
A substantial increase in computational throughput
using a simplified control structure and low hardware complexity was achieved,
together with low I/O and for low hardware cost~\cite{ChiperSwamyAhmad}.

An algorithm for computing the DHT using
an algebraic integers encoding scheme for
the coefficients is available in~\cite{Dimitrov1}.
An associate fully pipelined systolic architecture
with
$\mathcal{O}(N)$
throughput is available~\cite{Dimitrov1}.

A systolic-array processor for the prime-factor DHT
is given in~\cite{MeherSatapathyPanda}.
In~\cite{SundarBanerjee},
a systolic-array architecture for the fast computation of
the 2-D DHT of size $M \times N$ in $\mathcal{O}(M + N)$ time
using Givens rotors as processing elements is provided~\cite{SundarBanerjee}.
In critical applications,
fault detection of these VLSI circuits is also required.
A self-checking array architecture for
the radix-2 FHT transform is available in~\cite{TahirDlayNaguib}
where it is shown that algorithm-based error detection schemes for FHT
can be realized with reasonable hardware and time overheads.

\subsection{1-D and 2-D HNTT}
\label{section.math}

Despite the achievements on DHT implementation,
the DHT is inherently a transformation defined over the real numbers.
Thus,
it is naturally prone to truncation and round-off errors.
A truly error-free transform should work on a numerical framework
where the concept of error or approximation is inexistent.
Galois field theory offers an adequate algebraic formalism for
such framework~\cite{blahut1997fast}.

Let $\mathrm{GF}(p)$ be a Galois field of odd characteristic~$p$.
An integer $a$ is said to be a quadratic nonresidue of $p$
if the congruence $x^2 \equiv a \mod{p}$ has no solution~\cite[p.~68]{hardy1975introduction}.
If $p \equiv 3 \mod{4}$, then
the element $-1$ is a quadratic nonresidue.
Therefore,
the associated Gaussian integer field can be defined as
$\mathrm{GI}(p) = \{ a + j b : a, b \in \mathrm{GF}(p) \}$,
where
$j^2 \equiv -1 \mod{4}$.
An element $a+j b \in \mathrm{GI}(p)$ is said to be unimodular
if $a^2 + b^2 \equiv 1 \mod{p}$.
The finite field Hartley function is
analogous to its real field counterpart
and is
given by
\begin{align*}
\mathrm{cas}(i)
\triangleq
\cos(i) + \sin(i),
\quad
i = 0, 1, \ldots, N-1,
\end{align*}
where the finite field cosine and sine functions are
\begin{align*}
\cos(i) = \frac{\zeta^i + \zeta^{-i}}{2},
\quad
\sin(i) = \frac{\zeta^i - \zeta^{-i}}{2j},
\end{align*}
respectively,
and $\zeta \in \mathrm{GI}(p)$ is a fixed element of order~$N$.
In addition,
if
$\zeta$ is a unimodular element,
then
$\mathrm{cas}(i) \in \mathrm{GF}(p)$,
$i = 0, 1, \ldots, N-1$~\cite{cintra2009fragile}.

If
a unimodular $\zeta$ element is selected,
then
the 1-D HNTT relates input and output vectors,
$\mathbf{x} = \begin{bmatrix}x_0 & x_1 & \cdots & x_{N-1} \end{bmatrix}^T$
and
$\mathbf{X} = \begin{bmatrix}X_0 & X_1 & \cdots & X_{N-1} \end{bmatrix}^T$,
respectively,
according to the following pair of relations:
\begin{align*}
\mathbf{X} &= \mathbf{H}_N \cdot \mathbf{x}, \\
\mathbf{x} &= {N^{-1}\!\!\!\!\!\! \mod{p}} \cdot \mathbf{H}_N \cdot \mathbf{X},
\end{align*}
where $\mathbf{H}_N$ is the transformation matrix,
whose elements are given by
$[\mathbf{H}_N]_{i,k} = \mathrm{cas}(ik)$,
$i,k = 0, 1, \ldots, N-1$.
The unimodularity of $\zeta$ ensures that
if
$x_i \in \mathrm{GF}(p)$, for all~$i$,
then
$X_k \in \mathrm{GF}(p)$, for all~$k$~\cite{cintra2009fragile}.
Quantities $\zeta$, $p$, and $N$ are not independent.

Among several possible choices of $\zeta$, $p$, and $N$,
we adopted $\zeta = j$, $p=3$,
which implies $N=4$.
This combination of Galois field characteristic,
blocklength size and $\zeta$ element
under the HNTT formalism advances the existing techniques
in several ways.

Firstly,
whereas the method described in~\cite{tamori2002fragile}
employs the
standard Fourier-like NTT,
our approach employs the HNTT.
For this particular transform,
the above selected parameters
furnishes a particularly interesting
transformation matrix:
\begin{align*}
\mathbf{H}_4=
\begin{bmatrix}
1 & 1 & 1 & 1 \\
1 & 1 & 2 & 2 \\
1 & 2 & 1 & 2 \\
1 & 2 & 2 & 1
\end{bmatrix}
\equiv
\begin{bmatrix}
1 & \phantom{-}1 & \phantom{-}1 & \phantom{-}1 \\
1 & \phantom{-}1 & -1           & -1 \\
1 &           -1 & \phantom{-}1 & -1 \\
1 &           -1 &           -1 & \phantom{-}1
\end{bmatrix}
\hspace{-4mm}
\mod{3}
,
\end{align*}
since
$2\equiv -1 \mod{3}$.
In comparison,
the Fourier-like NTT approach
described in~\cite{tamori2002fragile}
requires a transformation matrix filled at best with powers of two.
Our design is computationally simpler:
it minimally requires only
additions and subtractions,
being multiplication-free in nature.

Secondly,
the HNTT can provide a symmetrical transform operation.
This is because
the forward transform matrix
is identical to
the inverse transform matrix.
Among the trigonometric based transforms,
the Hartley transform is unique in this aspect~\cite{bracewell1986hartley}.
This implies that
the hardware design for the forward transform
can be re-used for the inverse transform.

Thirdly,
in our mathematical setup,
even the usual scaling factor $N^{-1}$ present
in the inverse transformation could be eliminated.
This is possible due to the fact that,
for the chosen Galois field,
we have that $N^{-1}$ is congruent to one:
$N^{-1} \equiv 4^{-1}\equiv 1 \mod{3}$.

A possible drawback to this approach
would be the cumbersome 2-D HNTT framework,
inherited from the 2-D DHT formalism~\cite{bracewell1986hartley}.
Indeed, given a $4\times4$ matrix~$\mathbf{A}$,
its 2-D HNTT transform is furnished
by~\cite{cintra2009fragile}:
\begin{align*}
\frac{1}{2}
\left(
\mathbf{B} +
\mathbf{B}^{\text{(c)}} +
\mathbf{B}^{\text{(r)}} -
\mathbf{B}^{\text{(c,r)}}
\right)
,
\end{align*}
where
\begin{align*}
\mathbf{B} = \mathbf{H}_4 \cdot \mathbf{A} \cdot \mathbf{H}_4,
\end{align*}
and
$\mathbf{B}^{\text{(c)}}$,
$\mathbf{B}^{\text{(r)}}$,
and
$\mathbf{B}^{\text{(c,r)}}$
are built from the temporary matrix~$\mathbf{B}$.
Their elements are respectively given by
$t_{i,N-j\pmod{N}}$,
$t_{N-i\pmod{N},j}$, and
$t_{N-i\pmod{N},N-j\pmod{N}}$,
where $t_{i,j}$ are the elements of~$\mathbf{B}$,
for $i,j=0,\ldots,N-1$.
Above mathematical description
stems from the properties of the $\mathrm{cas}$ function~\cite[p.~21]{bracewell1986hartley}.
It
is excessively computationally complex,
being unattractive for the goals of this paper.

However,
for our purposes,
there is no point in calculation the actual 2-D HNTT.
In fact,
the lack of physical meaning of the NTT spectrum
allows us to consider
the 2-D \emph{special} Hartley transform,
whose computational complexity is lower.
Therefore,
we define the
2-D special HNTT of a $4\times4$ matrix~$\mathbf{A}$
as
simply equal to~$\mathbf{B}$.

\begin{figure*}
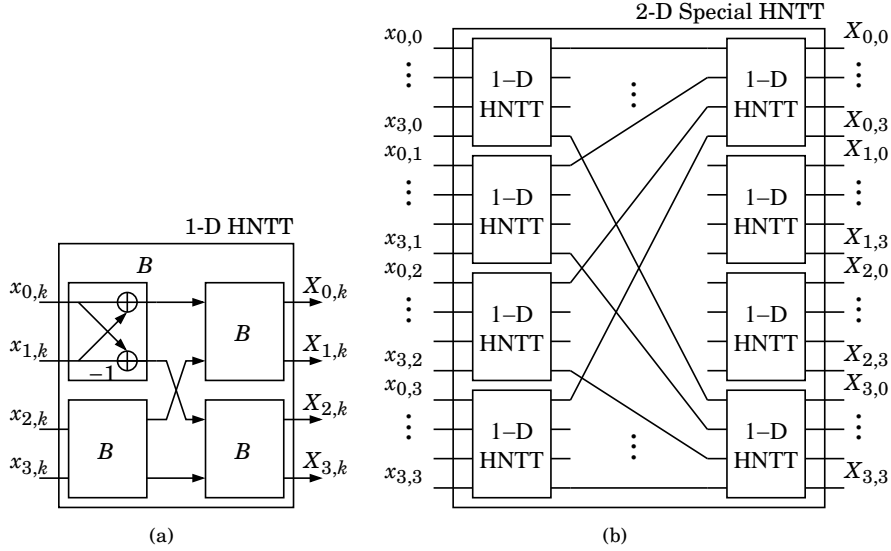
%
\centering
\subfigure[]{\input{fpga_ntt_fig1a.pstex_t}}
\quad\quad
\subfigure[]{\input{fpga_ntt_fig1b.pstex_t}}
\caption{%
Block diagram for
(a) the 4-point HNTT using butterflies ($B$) operating over $\mathrm{GF}(3)$
and
(b) the $4\times4$ special HNTT.}
\label{fig1}
\end{figure*}

The expression for $\mathbf{B}$
is the core operation of our proposed implementation.
It can be understood as
multiple applications
of
the 1-D HNTT to the columns of $\mathbf{A}$,
and then
to the rows of the resulting intermediate calculation.
This entire operation can be efficiently performed by means
of the fast algorithms as depicted in Fig.~\ref{fig1}(a)-(b).
By the end we obtained
a multiplication-free,
totally symmetric,
low complexity
2-D NTT.

\section{Watermarking Implementation and Results}
\label{section.implementation}

\subsection{Block-parallel systolic-array fragile watermarking architecture}

The proposed architecture employs
the 4-point HNTT based on modulo-3 arithmetic.
This configuration leads to relatively simple hardware.
Our architecture is massively-parallel,
fine-grain pipelined, and fully-systolic.
It is also modular, regular, and has good local interconnectivity,
making it well-suited for
application-specific integrated-circuit implementations.
We employ fully-parallel building-blocks for implementing the atomic arithmetic operation
$c \equiv a+b \pmod{3}$,
where $a$, $b$, and $c$ are 2-bits wide,
using 4-input look-up tables (LUTs) implemented on random-access memory (RAM).

The watermark embedding step is mathematically detailed in~\cite{tamori2002fragile}.
It
consists of the additive blockwise insertion
of a given watermark pattern
into the NTT domain of the input data blocks.
Input $4\times4$ 8-bit image blocks have their pixels
$x_{i,k}$, $0\leq i,k \leq 3$,
decomposed into
residue
$r_{i,k} \equiv x_{i,k} \pmod{3}$
and
divisible
$d_{i,k} = x_{i,k} - r_{i,k}$
parts.
This procedure is achieved by storing precomputed values of
$d_{i,k}$ in a LUT of depth 256.
The fully-pipelined architecture computes
the 2-D special HNTT (Fig.~\ref{fig1}(b))
of  incoming residues blocks at each new clock cycle,
resulting in number-theoretic transformed
residue blocks $R_{i,k}$.

Then
sixteen block-parallel modulo-3 adders
are used to insert
watermark pixels
$w_{i,k}$, $0\leq i,k \leq 3$
into NTT domain residue data $R_{i,k}$.
This operation is described according to the following expression:
\begin{align*}
R'_{i,k} \equiv R_{i,k} + w_{i,k} \mod{3},
\end{align*}
which is submitted
to
another instantiation of the 2-D special HNTT core.
This latter operation is intended to
inverse transform $R'_{i,k}$.

\begin{figure*}%
\centering
\input{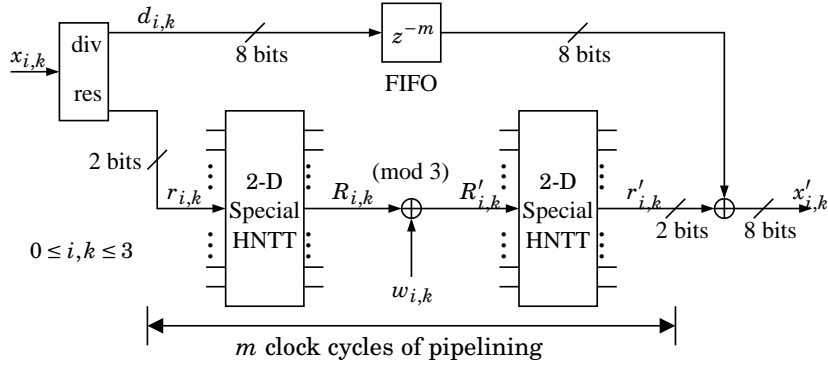}
\caption{%
Block diagram for
the watermark embedding scheme.}
\label{fig1.5}
\end{figure*}

Finally,
the resulting inverse transformed data $r'_{i,k}$ is
added to
the previously computed divisible part
employing
16 parallel 8-bit unsigned binary adders,
furnishing the watermarked data
\begin{align*}
x'_{i,k} = d_{i,k} + r'_{i,k}.
\end{align*}
The total throughout of the systolic-array is
one $4 \times 4$ block of watermarked samples per clock cycle
at a pipelining latency of $m$ cycles.
The design is shown in Fig.~\ref{fig1.5}.

The massively-parallel systolic-array proposed here
and shown in Fig.~\ref{fig1.5} is derived
using the NTT-based fragile watermark embedding algorithm detailed
in~\cite{tamori2002fragile}.
A direct-form realization is employed,
with feed-forward paths subject to fine-grain pipelining
for reduced critical path delay and
register retiming for lower dynamic power~\cite{KKparhi}.
The systolic-array is a one-to-one mapping of the algorithm.

\begin{figure*}
\centering
\epsfig{file=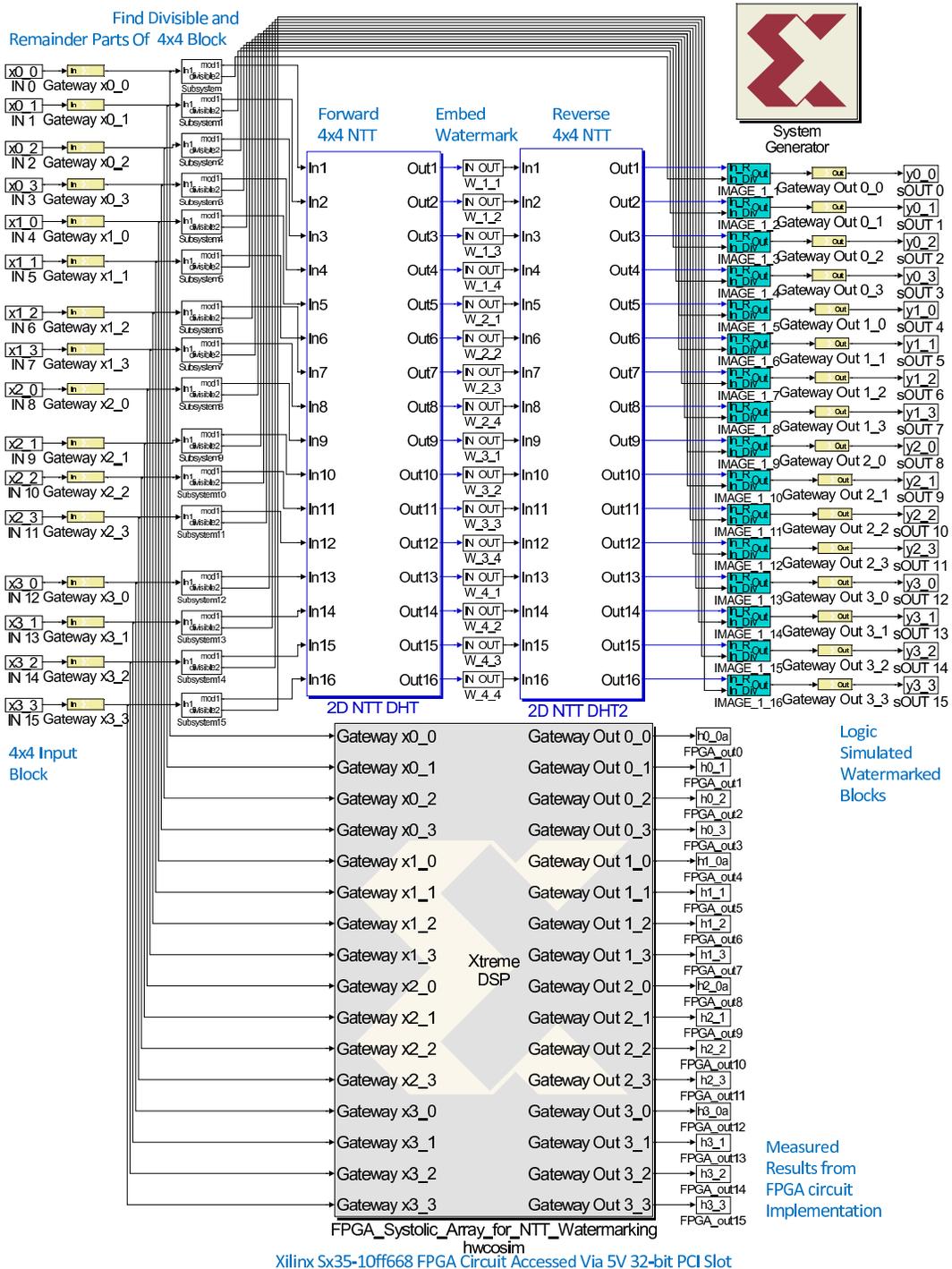,width=0.85\linewidth}
\caption{FPGA circuit design using Xilinx System Generator (XSG) with on-chip physical implementation.}
\label{fig3}
\end{figure*}

The IP core was physically implemented on a Xilinx Virtex-4 Sx35-10ff668 FPGA.
It consumed
2034 (out of 15360) slices,
3272 (out of 30720) logic fabric LUTs, and
160 (out of 192) FIFO16/RAM16 hardware blocks.
Post place-and-route timing analysis indicated a clock speed
greater than 100~MHz
for $m=89$ stages of internal pipelining.
Fig.~\ref{fig3} shows a screen shot of the final FPGA design and physical implementation,
which was completed using the bit-true cycle-accurate
model-based
design and implementation tool called Xilinx System Generator (XSG),
which functioned in association to Matlab and Simulink.
The XSG
could
automatically determine the glue-logic
required for the connection of the proposed systolic-array watermarking processor
to the personal computer memory space via the 32-bit PCI bus.

\subsection{Simulation}

The
on-chip
digital verification platform based on XSG allowed
stepped hardware cosimulation with a parallel software model in \textsc{Matlab}/Simulink.
This facilitated a direct comparison of simulated algorithm outputs
with actual measurements from the FPGA implementation under test.
An FPGA implementation of the proposed systolic-array for
the fragile watermark embedding was successfully verified
on-chip using hardware-in-the-loop cosimulation on
Xilinx XtremeDSP Development Kit-4 hardware prototyping system.

\begin{figure*}%
\centering
\subfigure[]{\epsfig{file=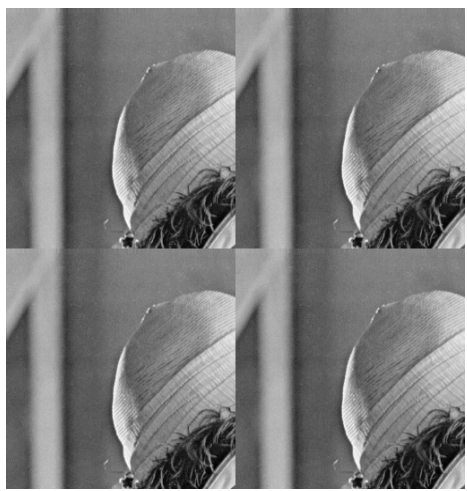,width=0.475\linewidth,height=0.475\linewidth}}

\subfigure[]{\epsfig{file=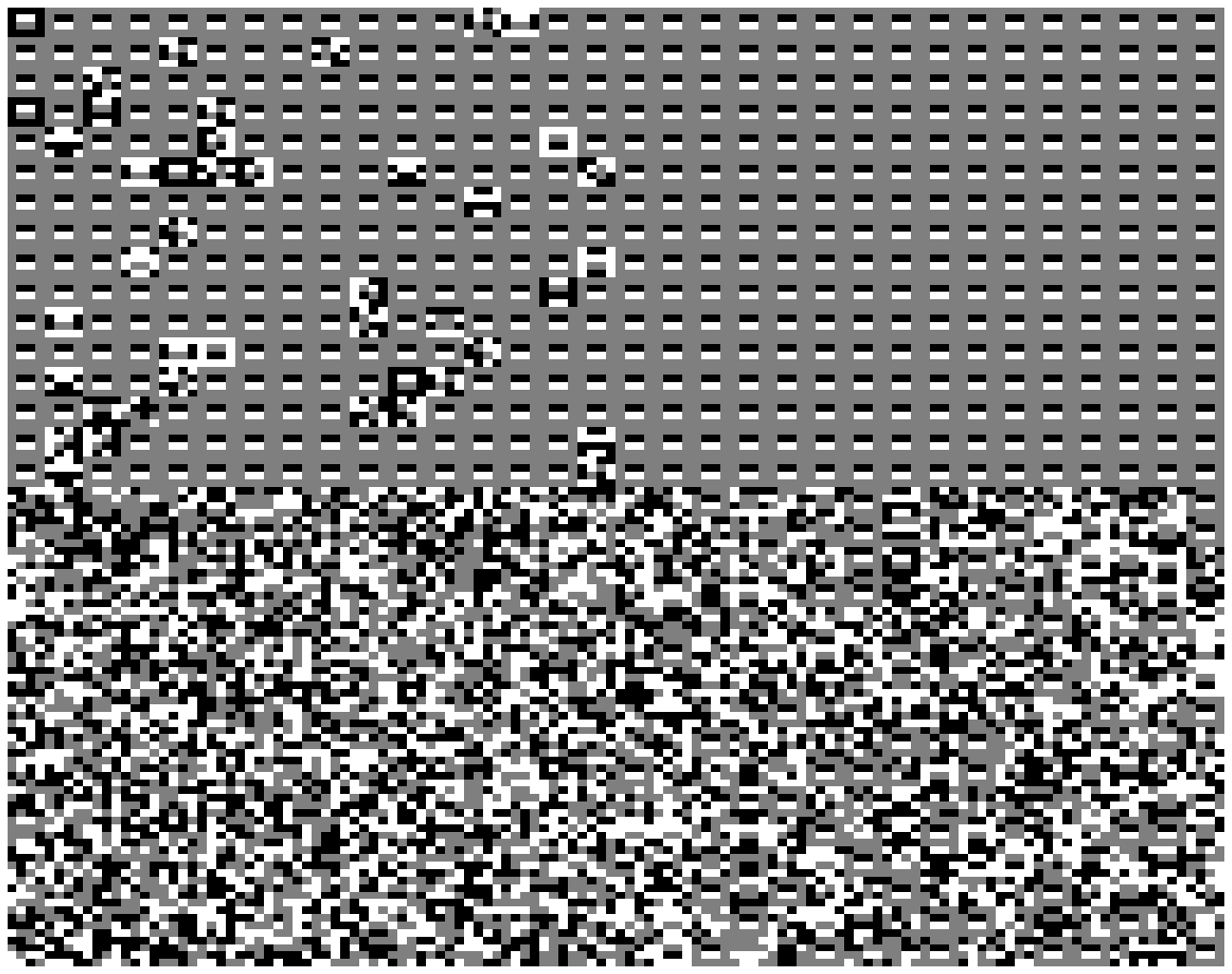,width=0.475\linewidth,height=0.475\linewidth}}
\caption{%
(a) Measured watermarked images from physical FPGA circuitry:
top right image was not tampered,
and remaining images were submitted to minimal perturbations (see text).
(b) Extracted watermarks for each case.}
\label{fig2}
\end{figure*}

Streamed $4\times4$ sized test data blocks from the Lena image
were submitted to the FPGA chip.
The resulting watermarked data
from the proposed core
was routed back to our computing environment via a PCI slot,
for analysis in \textsc{Matlab}.
Fig.~\ref{fig2}(a) shows intact and tampered versions of a
watermarked subimage of Lena portrait.

The following minimal image perturbations were considered:
(i) random changes of the least significant bit of each pixel with probability $10^{-2}$ (top left image);
(ii) JPEG compression and decompression with quality factor of 100 (bottom left image);
and
(iii) JPEG~2000 compression and decompression at 8 bits per pixel (bottom right image).
The employed watermark was the same in each case:
a regular pattern of pixels.

Watermarks are extracted by taking the modular difference between
the special HNTT transformed versions of the original image and the watermarked image.
This procedure is detailed in~\cite{tamori2002fragile}.
Obtained watermarks from subimages are shown homologously in Fig.~\ref{fig2}(b).
Clearly,
tampered subimages furnished severely damaged extracted watermarks.

The proposed 2-D fragile watermarking core delivered
one $4\times4$ block of watermarked results every clock cycle,
implying a video frame rate $95.3$~Hz at a resolution of $4096 \times 4096$,
for a clock of $100$~MHz.

\section{Conclusion}
\label{section.conclusions}

The
increasing demand for secure data storage and
tamper-free transmission is expected to lead to new applications
of high-speed server side hardware algorithms
for handling
large amounts of data and Internet traffic.
In this paper,
we proposed,
physically implemented,
and verified on chip
a novel systolic-array processor architecture for high-speed
efficient
fragile watermarking scheme for images.

The introduced circuitry is based on the
number-theoretic algorithm
introduced by Tamori~\emph{et al.}
in~\cite{tamori2002fragile}.
Our work also
introduced the number-theoretic version of the special DHT,
which was necessary for the transform-domain watermark embedding.
Finite field characteristic and unimodular element selection
were tailored to allow a low-complexity implementation
without complex arithmetic.

Moreover,
the proposed architecture is the only one of its kind available in the current literature
and is capable of embedding 100 million $4 \times 4$ image blocks per second,
in real-time, when implemented on a mid-capacity Xilinx Virtex-4 FPGA device.

\section*{Acknowledgments}
Financial support from
the College of Engineering, University of Akron, Ohio, USA;
CNPq and FACEPE, Brazil;
the Department of Foreign Affairs and International Trade, Canada;
and
the Natural Science and Engineering Research Council (NSERC), Canada,
is gratefully acknowledged.

{\small
\singlespacing
\bibliographystyle{siam}
\bibliography{ntt,r2_clean}
}

\end{document}